\documentclass[letter]{jpconf}
\usepackage{graphicx}
\usepackage[comma,numbers,sort&compress]{natbib}
\usepackage{iopams}
\usepackage{setstack}
\usepackage{amsfonts}
\usepackage{amssymb}
\usepackage{multirow}
\usepackage{color}
\usepackage{colortbl}
\usepackage{setspace}
\usepackage{verbatim}
\usepackage[hyphens]{url}
\usepackage[small]{caption}

\graphicspath{{figures/}}

\def\newblock{\hskip .11em plus .33em minus .07em}

\def\apj{Astrophys.\ J.}
\def\apjl{Astrophys.\ J. Lett.}
\def\apjs{Astrophys.\ J. Supp.\ Ser.}

\def\aap{Astron.\ Astrophys.}

\def\physrep{Phys.\ Rep.\ }
\def\mnras{Mon.\ Not.\ Roy.\ Astron.\ Soc.}

\def\prl{Phys.\ Rev.\ Lett.}
\def\prd{Phys.\ Rev.\ D.}
\def\prc{Phys.\ Rev.\ C.}

\def\apss{Astrophys.\ Space Sci.}

\newcommand{\code}[1]{\texttt{#1}}

\begin{document}
\title{Computational Models of Stellar Collapse and Core-Collapse Supernovae}

\author{Christian D Ott$^{1,2,3}$, Erik Schnetter$^{3,4}$, Adam
  Burrows$^{5}$, Eli Livne$^6$, Evan O'Connor$^{1}$, and Frank L\"offler$^{3}$}

\address{$^1$ TAPIR, Mailcode 350-17, California Institute of Technology, Pasadena, CA, USA}

\address{$^2$ Niels Bohr International Academy, Niels Bohr Institute, Copenhagen, Denmark}

\address{$^3$ Center for Computation \& Technology, Louisiana State University, Baton Rouge, LA}

\address{$^4$ Department of Physics \& Astronomy, Louisiana State University, Baton Rouge, LA, USA}

\address{$^5$ Department of Astrophysical Sciences, Princeton University, Princeton, NJ, USA}

\address{$^6$ Racah Institute of Physics, Hebrew University, Jerusalem, Israel}

\ead{cott@tapir.caltech.edu}

\begin{abstract}
Core-collapse supernovae are among Nature's most energetic
events. They mark the end of massive star evolution and pollute the
interstellar medium with the life-enabling ashes of thermonuclear
burning. Despite their importance for the evolution of
galaxies and life in the universe, the details of the core-collapse
supernova explosion mechanism remain in the dark and pose a daunting
computational challenge. We outline the multi-dimensional,
multi-scale, and multi-physics nature of the core-collapse supernova
problem and discuss computational strategies and requirements for its
solution.  Specifically, we highlight the axisymmetric (2D)
radiation-MHD code \code{VULCAN/2D} and present results obtained from
the first full-2D angle-dependent neutrino radiation-hydrodynamics
simulations of the post-core-bounce supernova evolution. We then go on
to discuss the new code \code{Zelmani} which is based on the
open-source HPC \code{Cactus} framework and provides a scalable AMR
approach for 3D fully general-relativistic modeling of stellar
collapse, core-collapse supernovae and black hole formation on current
and future massively-parallel HPC systems. We show \code{Zelmani}'s
scaling properties to more than 16,000 compute cores and discuss first
3D general-relativistic core-collapse results.
\end{abstract}

\section{Introduction}
Core-collapse supernova (CCSN) explosions are powered by the release
of gravitational energy in the collapse of a massive star's core to a
protoneutron star (PNS). 
While this general CCSN picture may be clear, its details
have evaded understanding despite many decades of concerted
theoretical and numerical effort.

When thermonuclear core burning ends, the core of a massive star (i.e., $M
\gtrsim 8-10$ {solar masses} $[M_\odot]$) is comprised of
iron-group nuclei (or O/Ne nuclei in the lowest-mass massive stars)
and supported against gravity's pull primarily by the degeneracy
pressure of relativistic electrons. Shell burning adds mass to this
core and eventually pushes it over its maximum supportable
mass. Gravitational collapse results and is accelerated by the capture
of electrons on free and bound protons and by the photodissociation of
heavy nuclei into alphas and nucleons. Core collapse continues until
the central, subsonically-collapsing region (the \emph{inner core} of
$\sim 0.5\,M_\odot$) reaches nuclear density. There, the strong
nuclear force kicks in, stiffening the nuclear equation of state (EOS)
and halting collapse, resulting in the rebound of the inner core into
the still collapsing outer core.  A hydrodynamic shock is formed in
this \emph{core bounce} and initially propagates outward in mass and
radius while losing energy to the dissociation of heavy infalling
nuclei and neutrinos that stream away from the postshock region. The
shock stalls within $\sim 10-20\,\mathrm{ms}$ after bounce and at a
radius of $\sim 100 - 200\,\mathrm{km}$. For a successful CCSN
explosion, it must be re-energized, otherwise continued accretion will
push the PNS over its mass limit. This results in a second phase of
gravitational collapse, leading to the formation of a black hole,
turning the stellar collapse event into a collapsar\footnote{By
  \emph{collapsar} we mean a stellar collapse event that does not lead
  to a supernova explosion. If the progenitor star has the needed
  angular momentum distribution, a Collapsar-type Gamma-Ray Burst may
  occur in a collapsar, see e.g.,~\cite{wb:06}.} \cite{heger:03}.

Finding and understanding the mechanism of shock revival has been the
key problem of core-collapse supernova theory in the past $\sim 45+$
years. Current theory and modeling suggests (see,
e.g., \cite{ott:09b,janka:07} and references therein) that there are
(at least) three ways to blow up massive stars: (1) the neutrino
mechanism, relying primarily on an imbalance between charged-current
neutrino heating and cooling in the immediate postshock regions in
combination with convection and the standing-accretion-shock
instability (SASI) (e.g., \cite{marek:09,scheck:08,murphy:08,bruenn:09}), (2)
the magnetorotational mechanism, based on bipolar jets
created by strong magnetic stresses in rapidly rotating cores (e.g.,
\cite{leblanc:70,bisno:76,burrows:07b}) and (3) the acoustic mechanism
recently proposed by
Burrows~et~al.~\cite{burrows:06,burrows:07a,ott:06prl}, 
which rests on the excitation of
non-radial pulsations in the PNS by accretion and turbulence and their
damping by strong sound waves that steepen into shocks, depositing
energy very efficiently in the postshock region. The acoustic mechanism
has not yet been confirmed by other groups and perturbative work suggests
that the pulsation amplitudes may be limited by a parametric instability
that transfers energy in faster damping daughter modes~\cite{weinberg:08}, but
this mechanism remains a compelling possiblity.

In the three potential explosion mechanisms, the breaking of spherical
symmetry is either fundamentally necessary or a key facilitating factor,
making \emph{multi-D modeling} a necessity. Furthermore, a CCSN
simulation must not only spatially resolve the steep gradients at the
PNS surface and the small to large scale turbulent eddies of overturn
(typical $\Delta x \sim \mathcal{O}(100\,\mathrm{m})$ or less for
magnetoturbulence~\cite{burrows:07b,cerda:07}), but the simulation
domain must also extend to many thousand kilometers to encompass
sufficient material to track a long-term postbounce accretion phase
without boundary effects affecting the evolution and to allow a
potential explosion to fully develop. Hence, \emph{multi-scale
  modeling} is required and may be implemented via fixed and/or
adaptive mesh refinement (FMR/AMR) or particle methods
(smoothed particle hydrodynamics [SPH], e.g.~\cite{rosswog:09}).
Furthermore, for a complete model of stellar collapse and the CCSN
postbounce phase, a broad spectrum of tightly-coupled physics must be
included -- ideally accurately, but at least approximately. This includes,
but is not necessarily limited to, (magneto)hydrodynamics (MHD),
general relativity (GR), nuclear physics (nuclear EOS and nuclear
reactions), neutrino radiation-transport, and neutrino-matter
interaction microphysics. 

The multi-D, multi-scale, and multi-physics nature of the CCSN problem
makes it complex and difficult to model and solve computationally. At
the same time, however, owing to its complexity, physically accurate
computational modeling, in combination with future detailed
observations of CCSN neutrinos and gravitational
waves\footnote{Gravitational waves are lowest order quadrupole
  waves. Hence, they are an intrinsically multi-D phenomenon and do not
exist in spherical symmetry. See \cite{thorne:87}
for a thorough introduction to gravitational wave theory.}, is
our only chance of solving it.

\vskip.2cm In this contribution to the Scientific Discovery through
Advanced Computing (SciDAC) Conference 2009, we discuss central
aspects of our broad computational approach to stellar collapse
and the CCSN problem and highlight recently obtained results. In
section~\ref{sec:vulcan}, we introduce the CCSN code \code{VULCAN/2D}
and present results from the first long-term full-2D momentum-space
angle-dependent radiation-hydrodynamics simulations of the postbounce
phase in CCSNe. We go on to describe in section~\ref{sec:3DGR} our full-GR
3D stellar collapse simulation package \code{Zelmani} which is based
on the \code{Cactus} computational framework~\cite{cactusweb1} and
designed for massively-parallel execution. \code{Zelmani} has already
been applied to simulations of rapidly-rotating 3D core collapse for
which we present results.
In Section~\ref{sec:conclusions}, we wrap up and present a forward-looking
summary.

\section{Angle-Dependent Neutrino Radiation-Hydrodynamic \code{VULCAN/2D} Simulations}
\label{sec:vulcan}

\begin{figure}
  \centering 
  \includegraphics[width=0.85\linewidth]{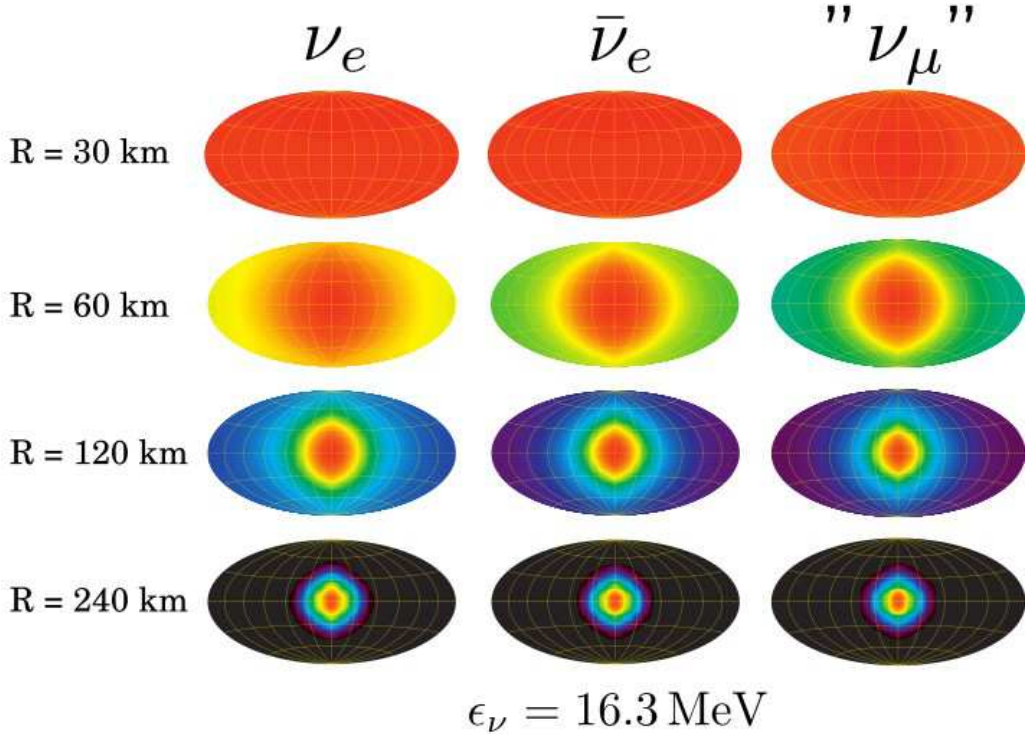}
  \caption{Hammer-type (smoothed) map projections of the normalized
    specific intensity $I_\nu(\vartheta,\varphi)/J_\nu$ (where $J_\nu$
    is the mean intensity) in model s20.nr at $160$~ms after
    bounce. The color map is logarithmic and each individual
    projection is set up to range from $\mathrm{max}(I_\nu/J_\nu)$
    (red) to $10^{-4}\,\mathrm{max}(I_\nu/J_\nu)$ (black).  Shown is
    the specific intensity of $\nu_e$, $\bar{\nu}_e$, and
    ``$\nu_\mu$'' neutrinos at $\epsilon_\nu = 16.3\,\mathrm{MeV}$ on
    the equator and at radii of $30$, $60$, $120$, and
    $240\,\mathrm{km}$.  The Hammer projection is set up in such a way
    that $\vartheta$ varies in the vertical from $0^\circ$ (top) to
    $180^\circ$ (bottom) and $\varphi$ varies horizontally from
    $-180^\circ$ (left) to $+180^\circ$ (right). Grid lines are drawn
    in $\vartheta$- and $\varphi$-intervals of 30$^\circ$. Note (a)
    that the neutrino radiation fields are isotropic at $R=30\,\mathrm{km}$,
    (b) the increasing forward-peaking of $I_\nu$ with increasing radius
    (and decreasing optical depth), and (c) that at any given radius
    $I_\nu$ of ``$\nu_\mu$'' is more forward-peaked than that of the
    $\bar{\nu}_e$ component, which, in turn, is always more
    forward-peaked than the $\nu_e$ component. This fact is a
    consequence of a transport mean-free path that varies with species
    (and energy; not shown here) and is smallest for the electron
    neutrinos.}
  \label{fig:psimap}
\end{figure}

\code{VULCAN/2D} is a general Newtonian axisymmetric (2D)
radiation-magnetohydrodynamics code described in
\cite{livne:93,livne:04,burrows:07a,ott:08} and extended and applied
to the stellar collapse and CCSN problems in a large number of studies
(e.g.,
\cite{ott:04,livne:04,livne:07,burrows:06,burrows:07a,ott:08,dessart:06pns,dessart:06aic,dessart:08a}).
\code{VULCAN/2D} implements the arbitrary Lagrangian-Eulerian (ALE)
technique with second-order TVD remap. The scheme is directionally
unsplit and allows for arbitrary grids. Here we use \code{VULCAN/2D}
in hydrodynamic mode which implements a finite-difference
representation of the Newtonian Euler equations with artificial
viscosity. \code{VULCAN/2D} allows for the use of general EOS tables
and for the present study we employ the finite-temperature nuclear
Shen EOS \citep{shen:98a} which is based on a relativistic mean-field
model for nuclear interactions and transitions to an ideal gas of
nuclei, nucleons, photons, and electrons at low densities.

\code{VULCAN/2D} implements neutrino transport in two different
multi-group (and multi-species) ways. The module implementing
time-implicit multi-group flux-limited diffusion (MGFLD) was described
in \cite{burrows:07a} and evolves the angle-averaged mean radiation
intensity $J_\nu$, the zeroth angular moment of the specific intensity
$I_\nu$, and uses the flux limiter of \cite{bruenn:85}. The
angle-dependent transport module implements the method of discrete
ordinates ($S_n$) \cite{livne:04, castor:04} in time-implicit fashion
at low optical depths and matches to MGFLD at optical depth $\tau
\gtrsim 2$ for accelerated conversion at high optical depths where the
radiation field is isotropic~\cite{ott:08}. Both transport solvers
employ the neutrino microphysics outlined in \cite{brt:06} (including
nucleon-nucleon bremsstrahlung, e.g., \cite{thomp:00}), use typically
$16$ energy groups logarithmically spaced from $2.5$ to
$220\,\mathrm{MeV}$, and consider $\nu_e$ and $\bar{\nu}_e$
individually, while lumping together $\nu_\mu$, $\bar{\nu}_\mu$,
$\nu_\tau$, and $\bar{\nu}_\tau$ into ``$\nu_\mu$.'' In both MGFLD and
$S_n$ we neglect neutrino energy-bin coupling (relevant in inelastic
scattering) and assume the slow-motion approximation to radiation
transport appropriate in the postbounce phase, neglecting
velocity-dependent terms of $O(v/c)$. As a consequence, neutrino
advection, Doppler shifts and aberration effects are not
considered. This greatly limits the computational complexity of the
problem, but its impact on the transport solution depends on the
particular choice of reference frame and was examined in
\cite{hb:07}. Around core bounce and neutrino breakout, during the
non-linear phase of the SASI hundreds of milliseconds after bounce,
and in the case of very rapid rotation, including $O(v/c)$ terms is
advisable. Full $O(v/c)$ Boltzmann transport with energy
redistribution will be addressed in the future.

In the $S_n$ solver, we discretize the angular radiation distribution
evenly in $\cos{\vartheta}$ from -1 to 1 and make the number of
$\varphi$-bins (running from $0$ to $\pi$, because of axial symmetry)
a function of $\cos{\vartheta}$ to tile the hemisphere more or less
uniformly in solid angle.  In our time-dependent $S_n$ runs, we employ
8 $\cos \vartheta$ bins, resulting in a total of 40 angular
zones. Steady-state radiation fields are computed
either with 8 $\cos \vartheta$ bins, 12 $\cos \vartheta$ bins (92 total
angular zones) or 16 $\cos \vartheta$ bins (162 total angular zones)
at each spatial grid point.

The computational costs for a \code{VULCAN/2D} simulation can be
estimated based on the single-zone update cost for one neutrino
group/species, the number of required updates, and the number of
zones, neutrino energy groups and species. For a typical MGFLD
simulation, $N_\mathrm{zones} = 40000$, $N_\mathrm{energy\, groups} =
16$, $N_\mathrm{species} = 3$ and the single-zone update cost
including hydro, gravity, and one group of MGFLD is $\sim 125\,
\mathrm{flop}$. Thus, a single timestep requires $\sim
240\,\mathrm{Gflop}$ and a typical simulation lasts for $\sim
1$~million timesteps, leaving us with a total $\mathrm{flop}$ count of
$\sim 250\,\mathrm{Pflop}$. In a $S_n$ simulation, the same single-zone
update $\mathrm{flop}$ count applies, but the number of zones is
scaled by a factor equaling the number of momentum-space angular zones
modified by an empirical correction factor of $\sim 0.1$ obtained
through timing measurements. Thus, a $S_8$ simulation is about $4$
times more expensive than a MGFLD run and requires $\sim
1000\,\mathrm{Pflop}$ to complete.

In order to complete a simulation in reasonable time, we parallelize
\code{VULCAN/2D} transparently and efficiently via MPI in neutrino
groups (energy/species). Hence, for $16 \times 3$ groups we obtain a
speed-up of almost $48$ compared with a single-core calculation, since
only scalars are communicated at the end of each timestep.  Node-local
OpenMP parallelism is an additional way for increasing performance
and is currently under consideration. Domain decomposition, however, is not a
viable option, since the communication overhead due primarily to the
relatively small number of zones in a 2D simulation would quickly
dominate over any performance gains.

Assuming a sustained performance of $1\,\mathrm{Gflop}/\mathrm{core}$,
a MGFLD calculation is completed in $\sim 60\, \mathrm{days}$, while
the $S_n$ calculations presented here have required
$90-120\,\mathrm{days}$ on 48 cores, but were run for only a
limited amount of physical postbounce time.

\subsection{Simulation Setup}

We consider here two models based on the $20$-$M_\odot$ supernova
progenitor of \cite{whw:02}.  Model s20.nr is mapped onto the
\code{VULCAN/2D} grid without rotation while we impose a precollapse
central angular velocity of $\pi\,\mathrm{rad \, s}^{-1}$ in model
s20.$\pi$ and decrease it slowly with distance from the rotation axis
according to the simple rotation law specified in \cite{ott:04}. This
results in an essentially rigidly-rotating PNS core with a period of
$\sim 2\,\mathrm{ms}$ and strong rotational deformation of the entire
PNS and the postshock region. We choose a \code{VULCAN/2D} grid setup
with a pseudo-Cartesian central region that smoothly transitions to a
spherical grid at $20\,\mathrm{km}$. This not only removes the
coordinate singularity at the origin of spherical grids, but also
liberates the PNS core and allows for larger hydrodynamic timesteps
\cite{burrows:06,burrows:07a}. In the angular direction we employ
$120$ zones and there are $30$ logarithmically-spaced radial
zones interior to the grid transition and $200$ (also
logarithmically-spaced) radial zones from $20\,\mathrm{km}$ to
$4000\,\mathrm{km}$.

Both s20.nr and s20.$\pi$ are run with MGFLD to $160\,\mathrm{ms}$
after core bounce. At this point, we solve for stationary-state
angle-dependent $S_n$ radiation fields, then evolve the simulations
with $S_8$ (see \cite{ott:08} for resolution tests) over postbounce
intervals of $340\,\mathrm{ms}$ and $390\,\mathrm{ms}$ for model
s20.nr and s20.$\pi$, respectively.  For comparison we also continue
the MGFLD variants.

\subsection{Results}
The CCSN problem is particularly challenging in its radiation
transport aspects, because the energy- and species-dependent neutrino
radiation fields transition from being completely isotropic inside the
PNS (where the diffusion limit applies) to being completely
forward-peaked at $\tau \ll 1$ (the free-streaming limit). The transition
between diffusion and free-streaming is handled in an approximate way
via the flux limiter in MGFLD (e.g., \cite{bruenn:85}), but
only angle-dependent transport can self-consistently and accurately
capture the gradual change of the radiation field whose degree of
forward-peakedness in the postshock heating region has an influence
on the neutrino heating efficiency~\cite{messer:98,ott:08}.
In Fig.~\ref{fig:psimap}, we present map projections of the angular
distribution of the specific neutrino radiation intensity $I_\nu$
as seen by an equatorial observer located at various radii. Shown
are the $I_\nu$ of $\nu_e$, $\bar{\nu}_e$, and ``$\nu_\mu$'' neutrinos
at a neutrino energy of $\epsilon_\nu = 16.3\,\mathrm{MeV}$. Inside
the PNS, at $R = 30\,\mathrm{km}$, all radiation fields are isotropic
and become increasingly forward peaked with radius, corresponding
to decreasing optical depth. At any given radius, ``$\nu_\mu$''s are more
forward-peaked than $\bar{\nu}_e$s which, in turn, are always more
forward-peaked than $\nu_e$s. This hierarchy is characteristic for the
postbounce phase of CCSNe and is a result of the species-dependent transport
mean-free path systematics in CCSN matter~(e.g., \cite{thompson:03}).

\begin{figure}
  \centering 
  \includegraphics[width=0.35\linewidth]{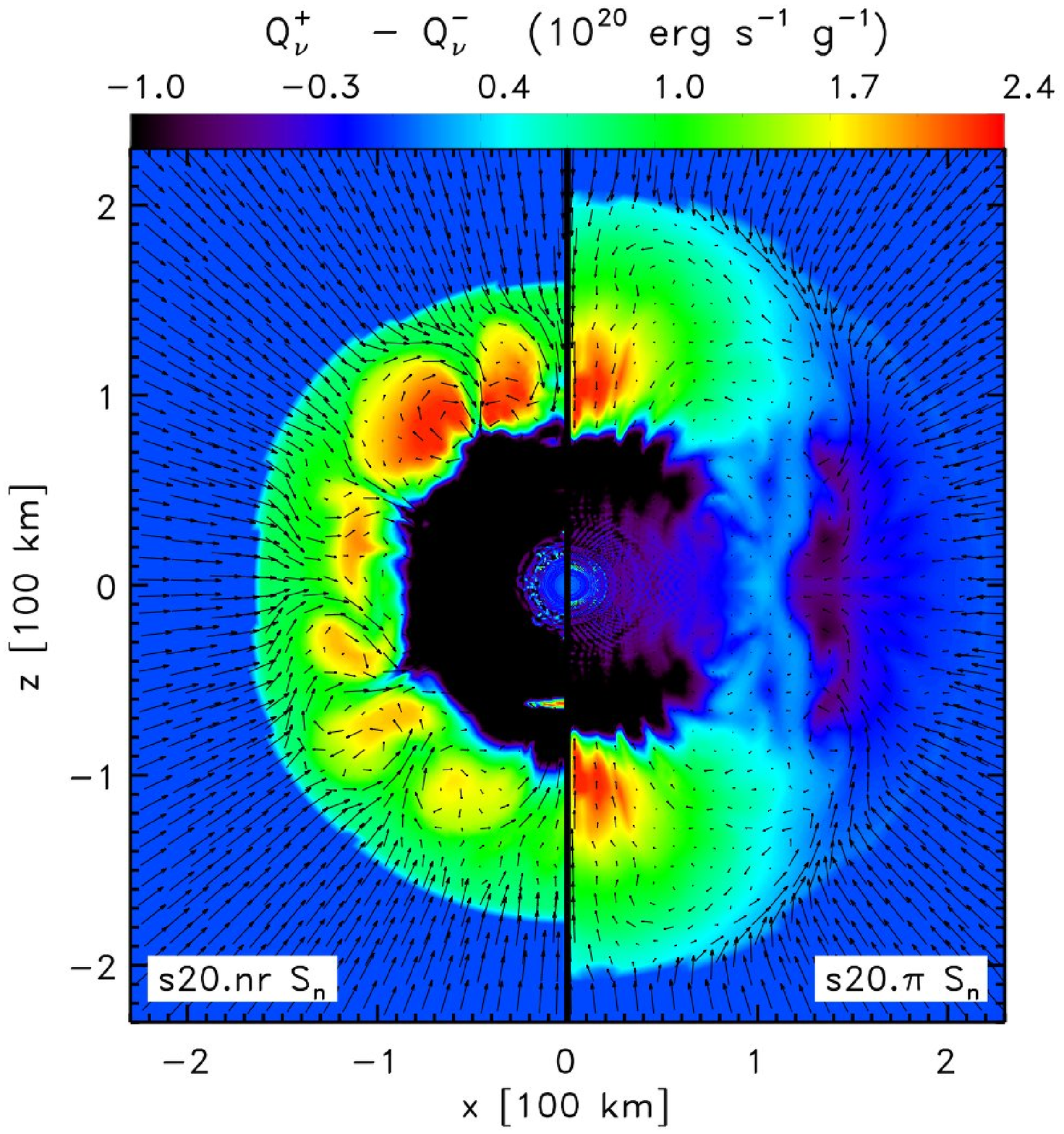}
  \includegraphics[width=0.35\linewidth]{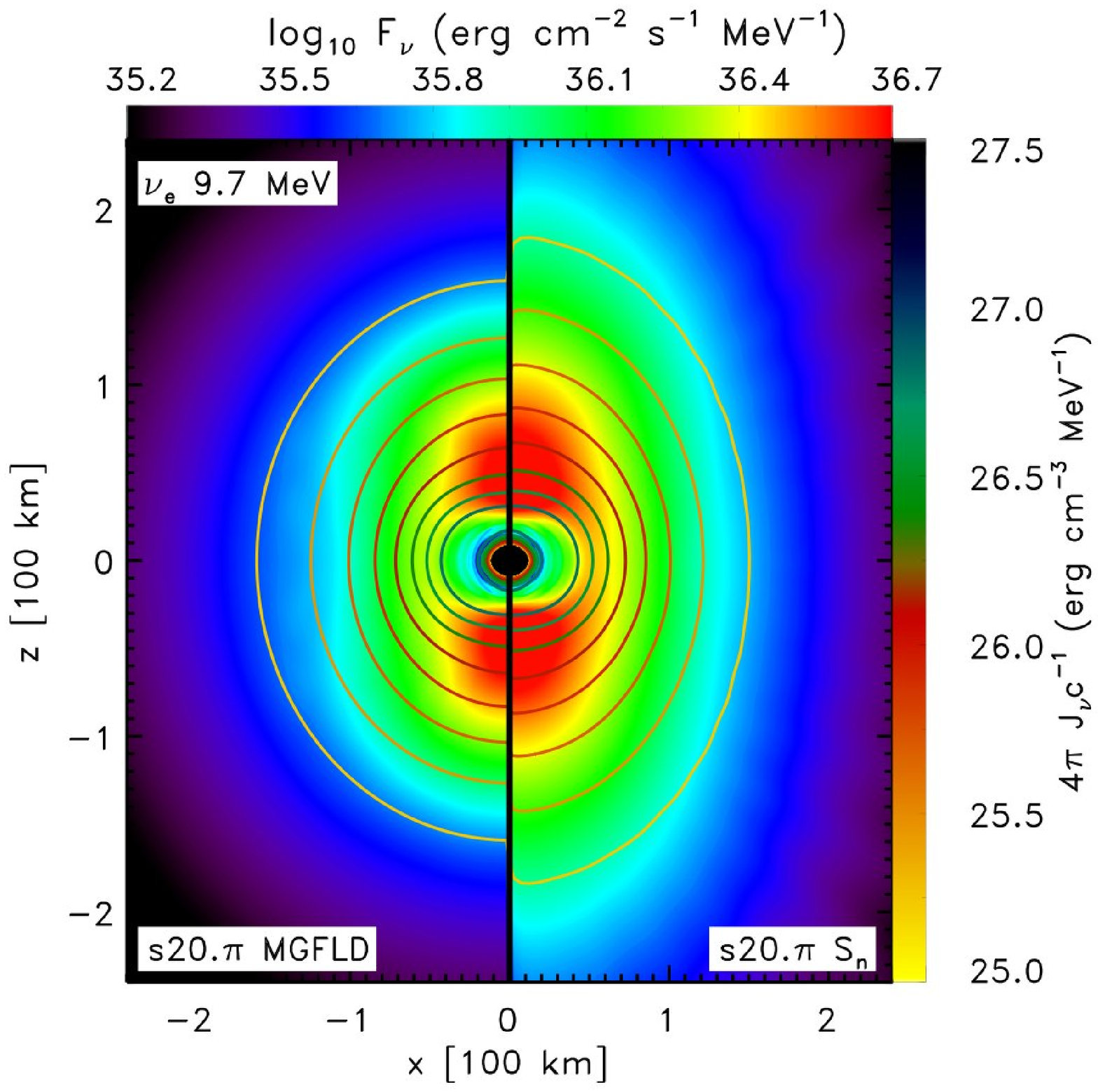}
  \caption{{\bf Left:} Comparison of the net gain (neutrino heating
    minus cooling) in the nonrotating model s20.nr and in the
    rapidly-rotating model s20.$\pi$ as obtained with 2D
    angle-dependent transport at $160\,\mathrm{ms}$ after bounce. Due
    to its rapid rotation, model s20.$\pi$ has a larger shock radius,
    enhanced heating near the axis, but due to a positive specific
    angular momentum gradient in the postshock region at low
    latitudes, also much less convective overturn. {\bf Right:}
    Comparison of the $\nu_e$ fluxes at $\epsilon_\nu =
    9.7\,\mathrm{MeV}$ in model s20.$\pi$ at $160\,\mathrm{ms}$ after
    bounce and as obtained with MGFLD (left section) and $S_n$ (right
    section). The radiation fields are oblate in the PNS core and
    deform to a prolate shape further out. Note that $S_n$ predicts a
    prolateness of the radiation field to much greater radii than
    MGFLD\@. The latter leads to nearly spherically symmetric radiation
    fields at radii greater than $\gtrsim\,$150--200~km. This
    result is largely independent of neutrino species and
    $\epsilon_\nu$.}
  \label{fig:gainflux}
\end{figure}

In the left panel of Fig.~\ref{fig:gainflux} we plot the specific net
gain, defined as neutrino heating $Q_\nu^+$ minus neutrino cooling
$Q_\nu^-$ and contrast the $S_n$ results for the nonrotating model s20.nr
(left half) with the rapidly rotating model s20.$\pi$ (right half) at
$160\,\mathrm{ms}$ after bounce.  In addition, we superpose velocity
vectors to visualize the flow of matter. The qualitative and
quantitative differences between the two models are large. In model
s20.nr, there is strong neutrino heating in the immediate postshock
region which drives strong convection. One also notes a slight
deformation of the shock away from spherical symmetry which is
indicative of the onset of the SASI\@. In model s20.$\pi$, on the
other hand, the neutrino heating occurs very asymmetrically and
primarily in polar regions where the neutrino flux is highest (see
also the discussion in \cite{walder:05,kotake:03nu,ott:08}). The shock
is more extended along the poles, but there is a pronounced equatorial
bulge with very little net heating. Convective overturn is confined to
polar regions due to (1) a large positive specific angular momentum
gradient at low latitudes, and (2) the absence of strong neutrino
heating in these regions \cite{fh:00,walder:05,ott:06spin}. In
addition, no signs of the SASI are apparent and the subsequent
evolution of model s20.$\pi$ suggests that rapid rotation delays and
modifies the SASI in axisymmetry. The situation may be different in
3D~(e.g., \cite{yamasaki:08,iwakami:08}).

The right panel of Fig.~\ref{fig:gainflux} visualizes the spectral
flux density of $\nu_e$ at $\epsilon_\nu = 9.7\,\mathrm{MeV}$ and at
$160\,\mathrm{ms}$ after bounce in the rapidly spinning model
s20.$\pi$ and contrasts the MGFLD result (left half) with that
obtained with $S_n$ (right half). In both variants, the radiation
field is oblate inside the PNS core, but quickly transitions to
a prolate shape further out. The density gradient in the polar regions
of the core is much steeper, allowing for much smaller neutrino sphere
radii ($R_\nu = R(\tau \approx 2/3)$) and resulting in a dramatic
enhancement of the polar neutrino flux~\cite{walder:05,kotake:03nu,
  ott:08}. The MGFLD result captures the overall systematics, but due
to the diffusive nature of the 2D MGFLD approach, the radiation field
asymmetry is smoothed out at low $\tau$ and becomes nearly spherical
at radii $\gtrsim 150\,\mathrm{km}$. The angle-dependent variant, on
the other hand, captures the true radiation-field asymmetry even at large
radii. In the case of model s20.$\pi$, this results in polar neutrino
heating that is locally larger by up to a factor of two than in the
MGFLD calculation. The radiation field asymmetry in the nonrotating
model s20.nr is much smaller and the radiation fields predicted 
by MGFLD and $S_n$ are more similar, while the $S_n$ calculation still
yields locally $\sim 10\%$ greater heating rates.

\begin{figure}
  \centering 
  \includegraphics[width=0.7\linewidth]{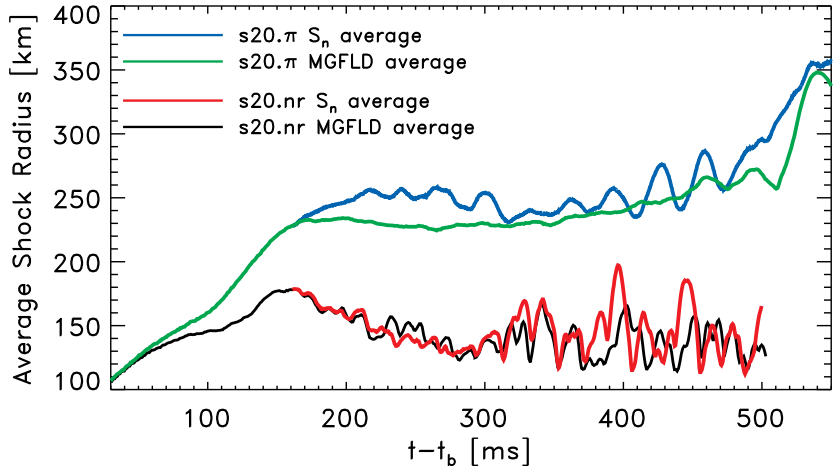}
  \caption{Average shock radius as a function of time after core
    bounce in MGFLD and $S_n$ variants of models s20.nr and
    s20.$\pi$. The large oscillations seen in the curves of the
    nonrotating model s20.nr are due to strong SASI oscillations and
    are somewhat more pronounced in the $S_n$ variant. The average
    shock radius of the rapidly rotating s20.$\pi$ settles at larger
    values than in the nonrotating case, but the growth of the SASI is
    slowed down by rapid rotation and only the $S_n$ model with its
    enhanced polar heating exhibits long-period SASI oscillations at
    intermediate postbounce times. See text and \cite{ott:08} for
  discussion.}
  \label{fig:shockradius}
\end{figure}

Figure~\ref{fig:shockradius} visualizes the dynamical effect of
angle-dependent neutrino transport on the postshock evolution of
CCSNe by comparing the time evolution of the angle-averaged shock
radii of the MGFLD and $S_n$ variants of models s20.nr and s20.$\pi$.
In the latter, switching to $S_n$ leads to a shock expansion primarily
in the polar regions where the increase in neutrino heating is
greatest. However, increased neutrino cooling from the now also
extended cooling region counteracts the shock expansion and lets it
settle at a new, somewhat higher average radius. The larger radius and
the increased local heating lead to an earlier onset of the
rotationally-modified SASI in the $S_n$ variant. This is reflected in
the earlier and more pronounced oscillations of the average shock
radius. However, towards the end of the postbounce time covered by
the simulations, the MGFLD variant's average shock radius catches
up and there is no large overall difference between
the dynamics seen in the  $S_n$ and MGFLD calculations of model
s20.$\pi$.

The dynamical evolutions in the $S_n$ and MGFLD variants of the
nonrotating model s20.nr are even more similar and the SASI shock
excursions in the two variants remain practically in phase for almost
$200\,\mathrm{ms}$. The $S_n$ calculation exhibits larger SASI
excursions at later times, but, as in the case of model s20.$\pi$,
there is no qualitative change between the MGFLD and $S_n$ postbounce
evolutions despite stronger neutrino heating (by up to $30\%$ of the
total heating rate at late times) in the latter.

\subsection{Discussion}
Neutrinos carry away $\sim 99\%$ of the gravitational energy of a
neutron star formed in stellar collapse. Their transport and their
interactions with matter are a central aspect of any CCSN model.
Here, we have presented and summarized results from our recent
\code{VULCAN/2D} simulations \cite{ott:08} that for the first time
addressed for a multi-D simulation the long-term dependence of the
postbounce dynamics of CCSNe on the neutrino transport technique
employed. Comparing MGFLD and angle-dependent $S_n$ transport, we find
that the former has difficulties in capturing physical radiation field
asymmetries and preserving them at low optical depth. The $S_n$
approach self-consistently evolves the radiation field from the
diffusion limit and isotropy to the free-streaming limit and
forward-peakedness. $S_n$ generally leads to locally stronger neutrino
heating, but the feedback in the supernova engine is sufficiently
strong to lead to an adjustment of the system to a new equilibrium
configuration without a true qualitative change when compared with
MGFLD\@. Thus, it appears unlikely that differences in neutrino
transport and/or interactions that result in by $10\% - 30\%$
increased heating (as seen in our simulations) can have a strong
impact on the CCSN dynamics. Larger effects seem necessary to turn the
CCSN dud into an explosion.

The work of the Garching group~\cite{buras:06b,marek:09} and of the
ORNL/FAU group~\cite{bruenn:09} suggest that the inclusion of a
general-relativistic monopole term in the otherwise Newtonian
potential in combination with a soft\footnote{These groups use the
  $K=180\,\mathrm{MeV}$ variant of the Lattimer-Swesty EOS which is
  too soft to support NSs with gravitational masses above $\sim
  1.7\,M_\odot$.}  EOS can increase the neutrino heating efficiency
due to hardened neutrino spectra and may lead to
explosion. Furthermore, \cite{murphy:08} have shown in a simplified
model with parametrized neutrino heating and cooling that explosions
are more easily obtained in 2D than in 1D\@. This result may very well
extend to 3D and supports and strengthens the CCSN community's motivation
to move towards 3D simulations.

\section{A New, Fully General-Relativistic Approach in 3D}
\label{sec:3DGR}

Today's technically most advanced and physically most complete stellar
collapse and long-term postbounce CCSN simulations are carried out in
axisymmetry (2D), use angle-dependent neutrino transport or MGFLD
either in full 2D~\cite{ott:08,burrows:06,burrows:07b, swesty:09} or
in a ray-by-ray
approximation~\cite{buras:06b,marek:09,bruenn:06,bruenn:09} and
implement Newtonian (magneto)hydrodynamics and gravity either in
purely Newtonian fashion
\cite{ott:08,burrows:06,burrows:07b,swesty:09} or with a GR monopole
term in an otherwise Newtonian gravitational
potential~\cite{buras:06b,marek:09,bruenn:06,bruenn:09}.  The current
state-of-the-art for axisymmetric GR core-collapse calculations, which
traditionally have focussed on estimates of the GW signal from
rotating core collapse and bounce, is set by \cite{dimmelmeier:08} who
performed conformally-flat\footnote{The conformal flatness condition
  (CFC) is an approximation to GR
  in which the radiative degrees of freedom have been
  suppressed~\cite{isenberg:08}.
  CFC is exact in spherical
  symmetry and is accurate to $\lesssim 5\%$ in the core-collapse
  scenario~\cite{cerda:05,ott:07prl,ott:07cqg}.}  calculations of the
collapse and early postbounce phase with a finite-temperature ($T$)
microphysical EOS and a simple deleptonization scheme
\cite{liebendoerfer:05fakenu} for the collapse phase in lieu of
neutrino transport.

Current published 3D simulations do not yet rival their 2D counterparts in
accuracy and physical
completeness. Iwakami~et~al.~\cite{iwakami:07,iwakami:08} investigated
the SASI in 3D with steady-state initial conditions, spherical
Newtonian gravity, a finite-$T$ microphysical EOS, parametrized
neutrino heating/cooling, a cut-out core, and a fixed high accretion
rate at the outer boundary.
The Basel group~\cite{scheidegger:08}, focussing on the GW signal
of stellar collapse, performed 3D calculations of the
collapse phase with a finite-$T$ microphysical EOS and the deleptonization
scheme of \cite{liebendoerfer:05fakenu} during collapse, but neglected
neutrino transport and heating/cooling in the postbounce phase.

More physically accurate simulations with better neutrino physics and
transport in the postbounce phase are required to address the
explosion mechanism. Several groups are in the process of implementing
codes with the necessary features (e.g.,
\cite{bruenn:06,bruenn:09,liebendoerfer:09}).

In the following, we present our approach to 3D CCSN modeling which
builds upon the tremendous recent progress in numerical relativity
(see, e.g., \cite{pretorius:07}) and
implements GR hydrodynamics and full GR curvature evolution in a
variant of the Arnowitt-Deser-Misner (ADM)~\cite{Arnowitt62} $3+1$
formalism. This allows us to not only more accurately follow the CCSN
hydrodynamics, but provides for the capability to form black holes
dynamically and in 3D in failing core-collapse supernovae -- something
that is impossible in Newtonian or pseudo-GR formulations. Our approach
makes heavy use of the \code{Cactus} computational
toolkit~\cite{cactusweb1} and is optimized
for execution on supercomputers, implementing AMR, domain decomposition,
parallel I/O, and hybrid MPI/OpenMP parallelism for improved scaling on
massively-parallel systems.

In Section~\ref{sec:infrastructure}, we introduce our GR curvature and
hydrodynamics formulation and discuss the computational
infrastructure and the various physics components of our approach, and
highlight parallel scaling results.  In Section~\ref{sec:grresults}, we
go on to discuss results obtained with our approach by
\cite{ott:06phd,ott:07prl,ott:07cqg} in the first set of GR
simulations of rotating core collapse in 3D\@.

\subsection{Computational Approach, Application Codes and Parallel Performance}
\label{sec:infrastructure}

We employ the open source software framework \code{Cactus}
\cite{Goodale02a, cactusweb1} designed for computational scientific
and engineering problems.  It has a modular structure and enables
scalable parallel computation across different architectures, as well
as collaborative code development between different research groups.

\code{Cactus} consists of a central part, called the \emph{flesh}, that
provides core routines, and of components, called \emph{thorns}.  The
flesh is independent of all thorns and provides the main program,
which parses input parameters and activates the appropriate thorns,
passing control to thorns as required.  By itself, the flesh does very
little science; to do any computational task the user must compile in
thorns and activate them at run time.  Parallelism, communication,
load balancing, memory management, and I/O are handled by a special
component, the \emph{driver}, which is not part of the flesh and
which can be transparently exchanged.  The flesh (and the driver) have
complete knowledge about the state of the application, allowing
inspection and introspection through generic APIs. 

\code{Cactus} runs on all current mainstream architectures.
Applications, developed on standard workstations or laptops, can be
seamlessly run on clusters or supercomputers.  \code{Cactus} provides
easy access to many cutting-edge software technologies being developed
in the academic research community, including the Globus Metacomputing
Toolkit \cite{Allen00d,globusweb}, HDF5 \cite{hdf5web}
parallel file I/O, the PETSc scientific
library \cite{petsc-home-page}, AMR,
multi-block methods \cite{Schnetter06a}, web
interfaces \cite{cactuswebserverweb}, and advanced visualization tools
(e.g.\ VisIt \cite{visitweb}).

The \code{Einstein Toolkit} \cite{ES-Schnetter2008n} is a set of
\code{Cactus} thorns providing infrastructure and basic functionality
for GR applications codes using the variables of the ADM formalism
\cite{Arnowitt62}. Within a simulation, the ADM variables are
employed for coupling GR curvature evolution with matter and radiation
variables and also serve in run-time analysis of the simulation results, 
such as e.g.\ evaluating
constraints, locating apparent horizons \cite{Thornburg95,
  Thornburg2003:AH-finding} or event horizons \cite{Diener03a}, or
calculating gravitational wave signals.

\subsubsection{GR Curvature Evolution.}

The Einstein equations are necessary for the correct description of
gravity in the strong-field regime. They are a set of ten coupled,
non-linear wave-type partial differential equations.  We solve these
equations using the \emph{BSSN} formulation (e.g.,
\cite{Alcubierre99d} and references therein).  This formulation,
similar to ADM, breaks up the four-dimensional spacetime into $3+1$
dimensions, three spatial and one time dimensions.  This leads to $25$
hyperbolic time evolution equations coupled to $9$ elliptic constraint
equations.

Of the $25$ evolution equations, $8$ are not specified by the Einstein
equations, but instead have to be chosen as gauge conditions to
determine the time evolution of the curvilinear coordinates of
spacetime.  We choose the so-called $1+\log$ slicing condition and the
$\Gamma$-driver shift condition \cite{Alcubierre02a,ott:06phd}, which
are standard gauge conditions used BSSN\@.  They ensure stable,
long-term time evolutions.  The $9$ constraint equations have to be
satisfied initially, requiring solving an elliptic system for setting
up initial data, and remain then satisfied under time evolution up to
within the discretization error.  We monitor the constraints during time
evolution and do not re-solve them.  The resulting equations for the
BSSN system and the gauge conditions can be time-evolved with standard
discretization methods.

We are using the \code{Kranc} package \cite{Husa:2004ip,
  krancweb} for automatic code generation for the BSSN formulation,
gauge conditions, and constraint equations.  \code{Kranc} is a
\code{Mathematica} package which generates \code{Cactus} thorns from
equations.  Starting from equations in \code{Mathematica} format which
specify a system of PDEs in abstract index notation, \code{Kranc}
discretizes the equations and generates a complete \code{Cactus} thorn
that evaluates these equations.  \code{Kranc}-generated thorns use all
relevant \code{Cactus} APIs for initial data setup, analysis, time
integration, and AMR\@.

Automatic code generation greatly reduces the time and effort
necessary to implement the BSSN equations, since these contain about
5,000 individual terms.  \code{Kranc} also allows us to experiment
with modifications to the formulation, e.g.\ to increase accuracy near
singularities, and with modifications to low-level implementation
details (loop blocking, vectorization) to achieve higher
efficiencies on modern computer architectures.

We have implemented the BSSN equations in an open-source \code{Cactus} thorn
arrangement called \code{McLachlan} \cite{ES-Brown2007b, mclachlanweb}
in the XiRel \cite{ES-Tao2008a, xirelweb} project.  \code{McLachlan}
is a full-featured BSSN solver able to simulate not only stellar
collapse, but also relativistic binary systems of black holes
or neutron stars \cite{ES-Brown2007b}.

\subsubsection{GR Hydrodynamics, Microphysics and Neutrinos}
On the GR hydrodynamics (GRHD) side, we implement the Valencia
formalism of GRHD \cite{Banyuls97,Font02c} which is based
on a dimensionally-split flux-conservative high-resolution
shock-capturing finite-volume approach. These methods are generally of
lower order than those used for the spacetime curvature evolution,
i.e., second-order in space and second-order in time. The main reason
for this is not only that the discretization of the GR hydrodynamics
problem becomes much more complicated with increasing order, but also
that all higher-order methods must drop back to first order near
discontinuities in the flow in order to prevent spurious oscillations
in the solution.  GR magnetohydrodynamics (GRMHD) is an extension to
GRHD and can be straightforwardly implemented within the same general
formalism~\cite{cerda:08,Giacomazzo:2007ti}.

The \code{Zelmani} CCSN package proper is a collection of
\code{Cactus} thorns implementing GRHD, a finite-temperature nuclear EOS,
neutrino leakage, and heating
\cite{ott:06phd,ott:07prl,ott:07cqg,oconnor:09}. We generally use the
name \code{Zelmani} synonymously to refer to the entire set of codes
involved in our 3D GR CCSNe simulations, including \code{McLachlan},
\code{Carpet}, \code{CactusEinstein}, and \code{Cactus}.  The GRHD
module is based on a modified version of the open-source \code{Whisky}
code~\cite{Baiotti04,whiskyweb} that allows for general,
finite-temperature EOS and neutrino-matter interactions. In our
simulations, we employ PPM reconstruction of variables at cell
interfaces and the approximate HLLE solver~\cite{Einfeldt88} for the
relativistic Riemann problem.

Fully relativistic neutrino radiation transport is a formidable
problem. Multi-D GR formulations exist~(e.g.,
\cite{cardall:03,zink:08}), but are still awaiting their first
implementations. While we are actively exploring various ways to
implement GR transport, we presently resort to the deleptonization
scheme of \cite{liebendoerfer:05fakenu} for collapse and employ
neutrino leakage (e.g., \cite{ruffert:96}) and parametrized heating in
the postbounce phase~\cite{oconnor:09,murphy:08}. Neutrino pressure
contributions are included via the approximation discussed in
\cite{liebendoerfer:05fakenu}.

\subsubsection{Adaptive Mesh Refinement and Parallel Scaling.}
\code{Carpet} \cite{Schnetter-etal-03b, Schnetter06a, carpetweb} is
our AMR driver for the \code{Cactus} framework.  \code{Carpet} acts as
a driver layer for \code{Cactus}, providing adaptive mesh refinement,
multi-patch capability, and efficient parallelization and I/O\@.  We
make both \code{Cactus} and \code{Carpet} publicly available as open
source, and both are also used by a number of other numerical
relativity and computational astrophysics groups.

\begin{figure}
  \includegraphics[width=0.49\linewidth]{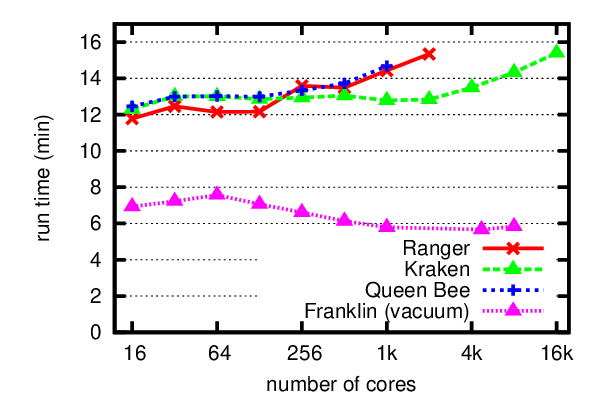}
  \includegraphics[width=0.49\linewidth]{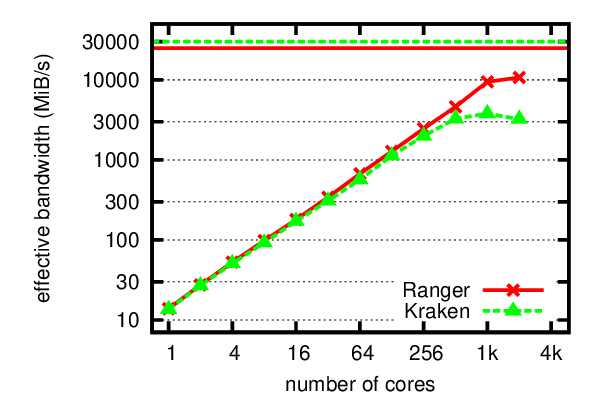}
  \caption{{\bf Left:} Weak scaling benchmark for spacetime curvature
    and GRHD evolution of a single neutron star (Ranger, Kraken, Queen
    Bee) or of a single black hole (Franklin) with nine levels of mesh
    refinement.  Ideal scaling would be an horizontal line in both
    cases.  \code{Cactus}/\code{Carpet} scales well up to more than
    16,000 cores.  Ranger and Kraken are the currently largest
    machines on the NSF TeraGrid, Queen Bee is the largest machine on
    LONI and Franklin is a large DoE/NERSC system.
    {\bf Right:} Cumulative I/O bandwidth for writing checkpoint
    files in parallel.  This measurement includes application
    overhead.  The theoretical peak bandwidth is shown as an horizontal
    line.  \code{Cactus}/\code{Carpet} is able to achieve a significant fraction of
    this.}
  \label{fig:scaling}
\end{figure}

\code{Carpet} provides spatial discretization based on highly
efficient block-structured, decomposed, logically Cartesian grids with
hybrid MPI/OpenMP \cite{mpiweb, openmpweb} parallelism.  \code{Carpet}
offers both AMR and multi-block capabilities, covering the domain with
sets of distorted, logically rectangular blocks of grids.  Time
integration is performed via the recursive Berger--Oliger AMR scheme
\cite{Berger84}, including subcycling in time.  As demonstrated in the
left panel of Fig.~\ref{fig:scaling}, \code{Carpet} presently scales
well to more than 16,000 cores with AMR on Leadership HPC
systems.

\code{Carpet} employs HDF5 \cite{hdf5web} for parallel, binary I/O,
which is also used for checkpointing and restarting.  The right panel
of Fig.~\ref{fig:scaling} depicts results of I/O benchmarks, comparing
the achieved I/O bandwidth to the theoretical peak bandwidth on
selected HPC systems. \code{Carpet} achieves a significant fraction of
this already on about 1,000 cores.

Although the speed and performance of high-end computers have
increased dramatically over the last decade, the ease of programming
such parallel computers has not progressed.  To address these issues
in computational modeling in general and in numerical relativity and
CCSN simulations in particular, we are developing the \emph{Alpaca
  tools} \cite{ES-Schnetter2007b, alpacaweb}.  In contrast to existing
debuggers and profilers, these tools work at the much higher level of
the physical equations and their discretizations and not at the level
of individual lines of code or variables.  The Alpaca tools are not
external to the application, but are built-in, so that they have
direct high-level access to information about the running application,
and can interact with the user on a correspondingly high level.

\subsection{Computational Costs, Simulations and First Results}
\label{sec:grresults}

A typical simulation setup to track collapse and postbounce CCSN
evolution with GR curvature evolution and GRHD uses a 9-level AMR grid
hierarchy with $\sim 400^3$ computational zones each, providing high
resolution near the center ($\delta x \lesssim 300\,\mathrm{m}$), while
encompassing the inner $\sim 5000\,\mathrm{km}$ of the dying star.
There are about $400$ 3D grid functions required for curvature and
GRHD which translates to a memory footprint of $\sim 2\,\mathrm{TB}$
(including inter-process buffers assuming 1024 processes).  A single
point update requires $\sim 50\,\mathrm{kflop}$ and $\sim
1\,\mathrm{million}$ fine grid updates are required, resulting in a
total cost of $\sim 1500\,\mathrm{Pflop}$ for a single simulation.  On
1024 compute cores and assuming a sustained performance of
$1\,\mathrm{Gflop/s}$ per core, such a simulation requires $\sim
17\,\mathrm{days}$ to complete. If radiation transport, even in
approximate ray-by-ray fashion, is included, the memory footprint
and the total number of required flops must be scaled by a factor
of $10-100$.

As a first application of our \code{Cactus}-based 3D GR \code{Zelmani}
CCSN code, we are considering the collapse and very early
postbounce phase of rapidly rotating iron cores with an emphasis on
the study of rotational multi-D dynamics leading to the emission of
gravitational waves (GWs). The latter may, in combination with neutrinos,
play an important future role as diagnostic tools for the CCSN
mechanism~\cite{ott:09,ott:09b} and in the case of rotating collapse
can provide information on the nuclear EOS, as well as on the
rotation rate of the inner core at bounce~\cite{dimmelmeier:08}.

We have carried out the first parameter study of rotating core
collapse in 3D GR, investigating the dependence of dynamics and GW
signal on progenitor stellar structure and precollapse rotational
setup, specified by the degree of differential rotation and the
initial central angular velocity. Since these simulations were run
with an earlier version of \code{Zelmani}, postbounce deleptonization
was neglected. The results of this study were extensively discussed in
\cite{ott:06phd,ott:07prl,ott:07cqg} and we highlight a number of the
findings in the following.

\begin{figure}
  \centering 
  \hspace*{-.5cm}
  \includegraphics[width=0.52\linewidth]{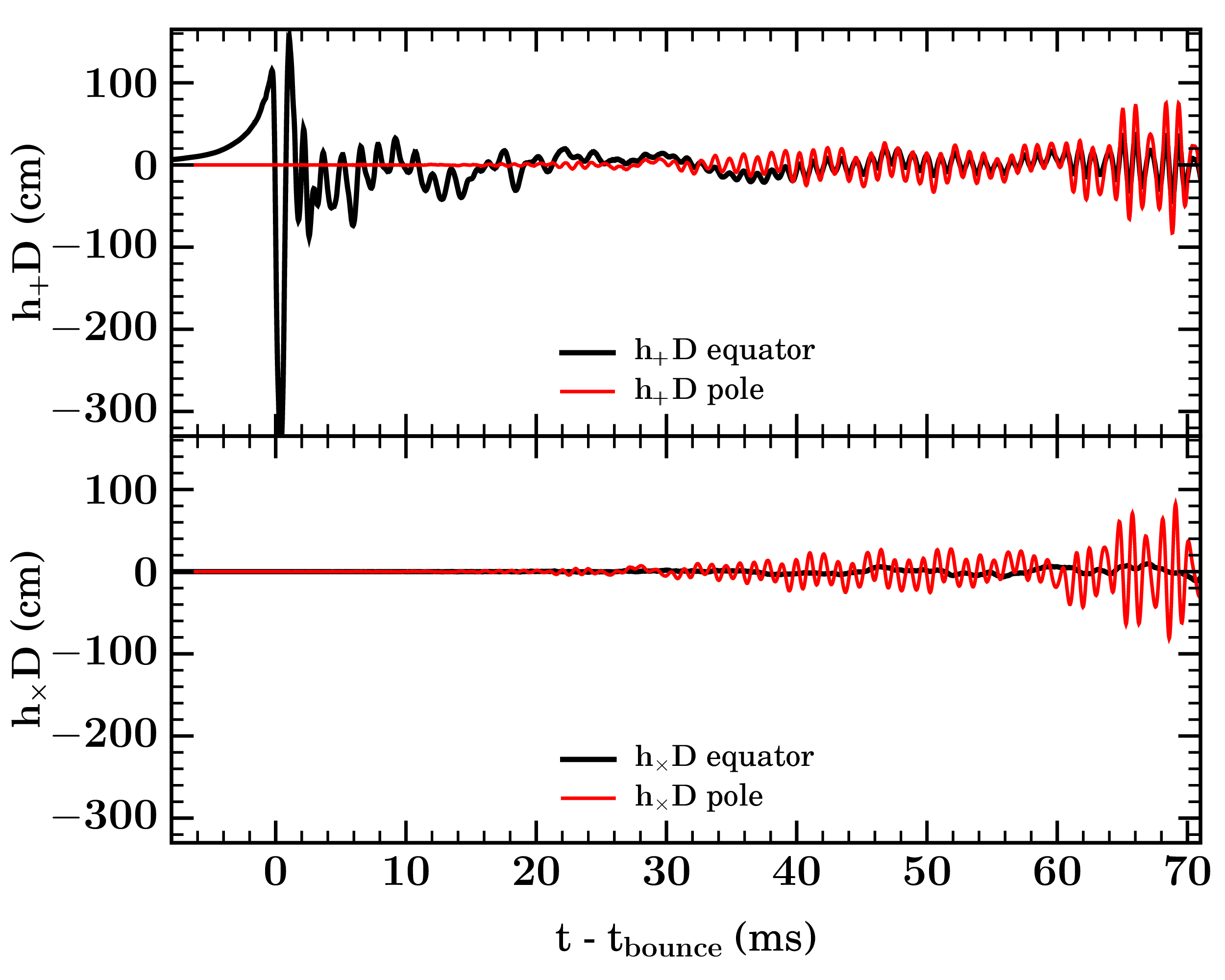}\hspace*{-.3cm}
  \includegraphics[width=0.5\linewidth]{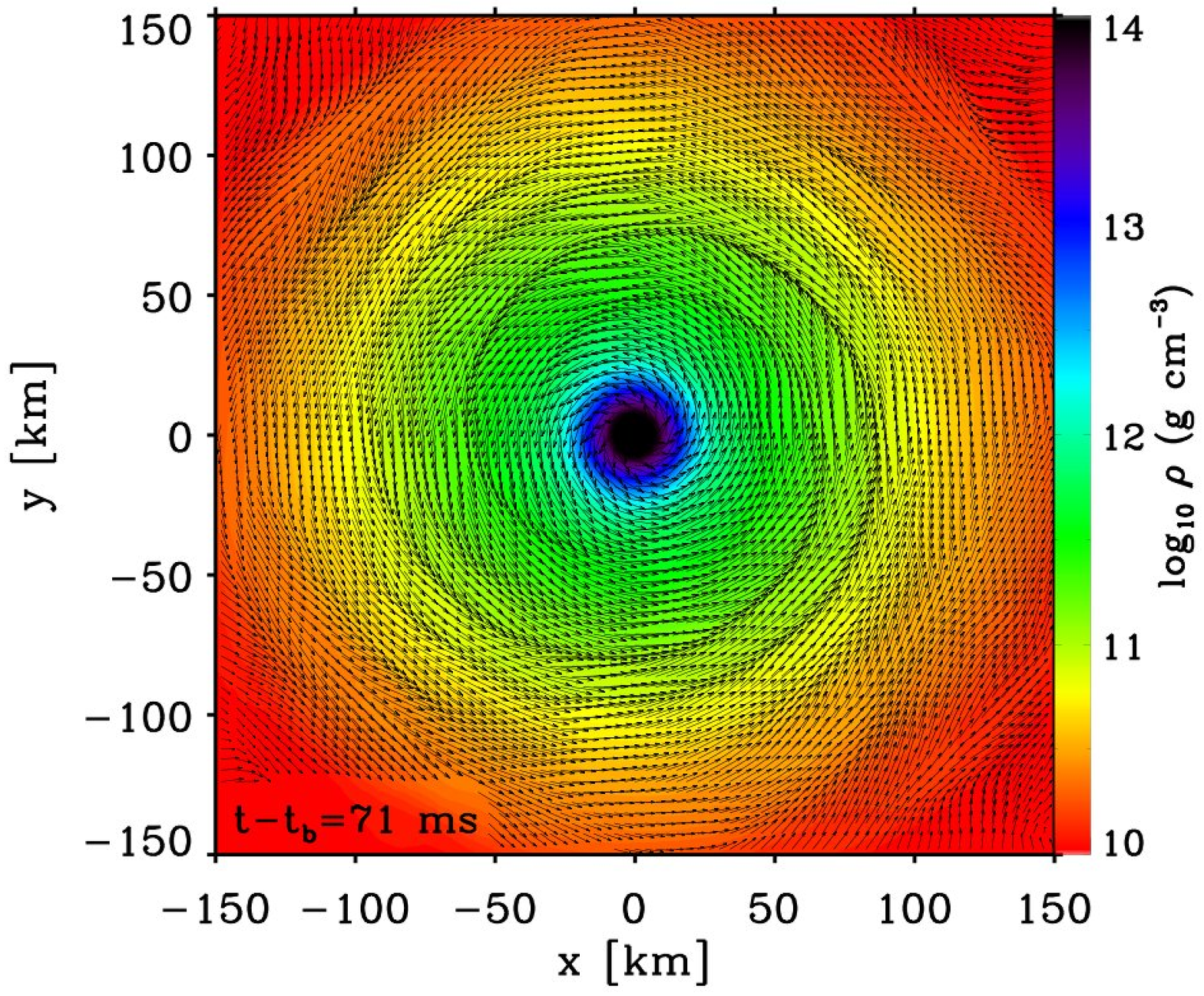}
  \hspace*{-.5cm}
  \caption{{\bf Left:} Gravitational wave (GW)
    signals of the $+$ polarization (top, $h_+\, D$, where $D$ is the
    source distance) and of the $\times$ polarization (bottom,
    $h_\times D$), emitted by model E20 as a function of postbounce
    time and as seen by polar (red lines) and equatorial (black lines)
    observers. Collapse and bounce are axisymmetric (only $h_+$ and
    emission only away from the symmetry axis) while at postbounce times
    nonaxisymmetric dynamics and GW emission develop.
    {\bf Right:} Density colormap showing the inner
    $150$x$150$~km of the equatorial plane ($z=0$) at
    $71\,\mathrm{ms}$ after bounce in the 3D GR model E20A\@. Velocity
    vectors are superposed and mark spiral density waves that develop
    due to a nonaxisymmetric instability in the PNS and propagate
    through the postshock region.  }
  \label{fig:E20A}
\end{figure}

In the left panel of Fig.~\ref{fig:E20A} we plot the two 
GW polarizations $h_+$ and $h_\times$, both scaled by distance $D$
to the source and in units of centimeters, as extracted from the
multi-D dynamics in our model E20A that is based on the rotating
$20$-$M_\odot$ progenitor star of \cite{heger:00}. Its precollapse
central angular velocity is $\sim 3.1\,\mathrm{rad\,s}^{-1}$ and its
precollapse rotation rate $\beta = T/|W|$ is $\sim 0.004$. The
corresponding values early after bounce are $\sim
4\,\mathrm{rad\,ms}^{-1}$ and $\sim 0.09$. The simulation is run
in 3D from the onset of collapse to $\sim 70\,\mathrm{ms}$ after
bounce.

We find that model E20A, like all other considered models, stays
axisymmetric through collapse, core bounce, and the early postbounce
phase. At bounce, the large deceleration of the inner core's infall
leads to a large negative spike in the waveform clearly
visible in $h_+ D$ as seen by an equatorial observer and shown in the
left panel of Fig.~\ref{fig:E20A}. Due to the axisymmetry of the
rotating collapse dynamics, GWs are emitted only in one polarization
(linear polarization; here, due to the choice of source orientation,
in the $+$ polarization) and only away from the symmetry axis of the
system. After bounce and on a timescale of $\gtrsim
20-40\,\mathrm{ms}$ nonaxisymmetric dynamics develops in the PNS and is
likely due to a corotation-type instability in which an
azimuthal mode picks up power from the axisymmetric background
rotation at the point where its mode pattern speed is in corotation
with the fluid~(see, e.g., \cite{watts:05}). The quadrupole components
($\ell=2,m=2$) of the nonaxisymmetric dynamics are reflected in the
late-time GW signal as a quasi-periodic, elliptically polarized
signal, strongly correlated in $h_+$ and $h_\times$ and emitted at
twice the $m=2$ pattern speed. This nonaxisymmetric GW emission is
smaller in amplitude than the burst associated with core bounce, but,
due to its longer duration, the total energy emission is greater and
the emission's narrowband nature favors detection by GW
observatories such as LIGO~\cite{LIGO,ott:09}.

Corotation instabilities are well known from studies of astrophysical
disks and belong to a class of dynamical shear instabilities that draw
from the shear energy stored in differential rotation and transport
angular momentum outward in spiral waves. Differential rotation is
abundant in the postshock region~\cite{ott:06phd,ott:06spin} and the
spiral waves are apparent in the right panel of Fig.~\ref{fig:E20A}
that depicts the density distribution with superposed fluid velocity
vectors in the equatorial plane of model E20A at $\sim
70\,\mathrm{ms}$ after bounce.

\section{Summary and Outlook}
\label{sec:conclusions}

The challenging complexity and non-linearity of the CCSN problem and
current technical and computational limitations call for a broad
computational program with multiple modeling approaches.  In this
contribution to SciDAC 2009, we introduced two CCSN simulation
programs and discussed their recent results. The first,
\code{VULCAN/2D}, a full CCSN code with an implementation of the
angle-dependent and MGFLD radiation-MHD equations, is capable of
studying all presently discussed CCSN mechanisms, but is limited to
axisymmetry and Newtonian gravity and dynamics. Being based on proven
legacy technology and despite not employing the massively-parallel
domain-decomposition paradigm, \code{VULCAN/2D} runs very efficiently
on a modest number of compute cores and yields a high science output
per flop (e.g.,
\cite{ott:04,livne:04,walder:05,burrows:06,burrows:07a,burrows:07b,dessart:06pns,dessart:06aic,ott:08}). This
includes the first long-term true multi-D angle-dependent neutrino
transport CCSN calculations highlighted in this article. The result of
this study is an example of \emph{Mazurek's law}\footnote{Mazurek's
  law originated in the context of stellar collapse at Stony Brook
  University in the 1980's when Ted Mazurek was there. It is now used
  to generally refer to the strong feedback in a complicated
  astrophysical situation which dampens the effect of a change in any
  single parameter~\cite{burrows:01,lattimer:09}.}. Applied to the
present situation, it states that in the tightly-coupled CCSN
phenomenon, even a rather significant ($\gtrsim 10\%-30\%$) change of
the postbounce conditions, in this case, of the neutrino heating rate,
is absorbed by the strong feedback between radiation, hydrodynamics,
EOS, and gravity and no qualitative change results.

The frontier of CCSN modeling is clearly 3D and the hope is that the
additional degree of freedom and the more accurate representation of
(turbulent) convection and SASI help produce successful and powerful
explosions whose asymptotic energies match observations.  We approach
the computationally and technically challenging step to 3D with the
code \code{Zelmani} that is based on the open-source \code{Cactus}
framework and uses scalable AMR via the \code{Carpet} driver module
that implements state-of-the-art HPC paradigms, including hybrid
MPI/OpenMP parallelism for modern massively-parallel multi-core
architectures. \code{Zelmani} is set apart from other 3D codes (e.g.,
\cite{scheidegger:08,bruenn:06,bruenn:09}) by its full GR nature,
evolving Einstein's equations from one spatial $3$-hypersurface to the
next and treating the dynamics fully relativistically instead of
re-solving a Newtonian or pseudo-relativistic Poisson equation at
every timestep coupled to Newtonian hydrodynamics. Dynamical spacetime
evolution not only enables us to study the impact of GR on the CCSN
evolution, but also allows us to investigate the dynamical formation
of black holes in failing core-collapse supernovae. Such
\emph{collapsars} are considered as likely candidates for the central
engines of gamma-ray bursts (GRBs; e.g., \cite{wb:06}), but their
dynamical formation has not been modeled and remains to be understood.

While still under development towards a full GR radiation-MHD CCSN
code, \code{Zelmani} has already been applied to 3D GR hydrodynamic
studies of rotating stellar collapse, leading to an improved
understanding of the gravitational wave signature of
CCSNe~\cite{ott:06phd,ott:07cqg,ott:07prl}.  In its completed form,
\code{Zelmani} will allow us to take full advantage of petaflop
supercomputers to comprehensively address the CCSN problem and the
CCSN-GRB relationship.

\ack It is a pleasure to acknowledge help from and stimulating
conversations with J.~Murphy, L.~Dessart, E.~Abdikamalov, G.~Allen,
D.~Arnett, W.~Benger, S.~Bruenn, P.~Diener, H.~Dimmelmeier, I.~Hawke,
I.~Hinder, K.~Kotake, C.~Meakin, B.~Messer, A.~Mezzacappa,
J.~Nordhaus, L.~Rezzolla, B.~Schutz, E.~Seidel, S.~Su, J.~Tohline, and
S.~Woosley. The \code{Cactus}-based part of this work would not have
been possible without the invaluable help of our late colleague Thomas
Radke whom we miss dearly both personally and professionally.  CDO is
supported by a Sherman Fairchild Prize Fellowship at Caltech and by an
Otto Hahn Prize of the Max Planck Society.  AB is partially
supported by the Scientific Discovery through Advanced Computing
(SciDAC) program of the US Department of Energy under grant number
DE-FC02-06ER41452. EO is supported by a
Natural Sciences and Engineering Research Council of Canada
postgraduate scholarship. The development of \code{Cactus}, the
Einstein Toolkit, and the performance tools is supported by the NSF
grants \emph{XiRel} (no.\ 0701566), \emph{Alpaca} (no.\ 0721915), and
\emph{Blue Waters} (no.\ 0725070).
Simulations and benchmarks were performed on Queen Bee at LONI under
allocation \texttt{loni\_numrel03}, and on Kraken at NICS and Ranger at TACC
under the NSF TeraGrid allocation TG-MCA02N014. Additional
computations were performed at the National Energy Research Scientific
Computing Center (NERSC), which is supported by the Office of Science
of the US Department of Energy under contract DE-AC03-76SF00098.

{\scriptsize 
\providecommand{\newblock}{}

}

\end{document}